\documentclass[11pt,a4paper]{article}
\usepackage[hyperref]{acl2017}
\usepackage{times}
\usepackage{latexsym}
\usepackage{amsmath}
\usepackage{mathtools}
\DeclarePairedDelimiter\ceil{\lceil}{\rceil}

\usepackage{url}
\usepackage{subfig}

\aclfinalcopy % Uncomment this line for the final submission
\title{A Survey of Community Question Answering}

\author{Barun Patra \\
  Indian Institute of Technology Delhi \\  
  {\tt cs1130773@iitd.ac.in} \\
}
\date{}

\begin{document}
\maketitle
\begin{abstract}
With the advent of numerous community forums, tasks associated with the same have gained  importance in the recent past. With the influx of new questions every day on these forums, the issues of identifying methods to find answers to said questions, or even trying to detect duplicate questions, are of practical importance and are challenging in their own right. This paper aims at surveying some of the aforementioned issues, and methods proposed for tackling the same.
\end{abstract}
%% ------START OF INTRODUCTION -------------
\section{Introduction} \label{sec:Introduction}
Community Question Answering has seen a spectacular increase in popularity in the recent past. With the advent and popularity of sites like Yahoo! Answers\footnote{https://answers.yahoo.com}, Cross Validated\footnote{http://stats.stackexchange.com/}, Stack Overflow\footnote{https://www.stackoverflow.com}, Quora\footnote{https://www.quora.com/}, more and more people now use these web forums to get answers to their questions. These forums give people the ability to post their queries online, and have multiple experts across the world answer them, while being able to provide their opinions or expertise to help other users, a quality that encourages more participation and consequently has lead to their popularity.     \\
This survey aims at discussing some of the challenges that accompany such community forums, and the way they have been approached. Section 2 defines the attributes of community question answering and contrasts it with the traditional question answering task. Section 3 defines some of the tasks seen in this domain and investigates the methods used to solve them. Section 4 defines the experimental setting and the datasets used by the methods mentioned in the previous section for evaluation purposes. Section 5 then summarizes the performance of various methods. Section 6 provides a general discussion of the results and finally, section 7 concludes this survey.

%% -----END OF INTRODUCTION ---------------
%% -----START OF cQA vs QA ----------------
\section{Community QA vs QA} \label{sec:DiffFromQA}
A community forum generally involves the following:
\begin{itemize}
\item The asker posts a query, which, after being checked for inappropriate content (by moderators) or sometimes duplication, is posted and is visible to other users for answering.
\item The other users interact in two ways :
\begin{itemize}
\item By posting relevant (or irrelevant) answers, based on their opinions/ expertise.
\item By upvoting or downvoting answers by other users, based on the validity, significance and content of the responses.
\end{itemize}
\item Some Community QA sites also allow the users to interact with the question by downvoting/ upvoting the question itself or by simply commenting on the question (asking for other details, or pointing the asker to other relevant questions.)
\item Finally, if the asker is satisfied, s/he may mark the best answer (either chosen by the maximum number of upvotes, or by the asker himself/ herself), and the question may be archived.
\end{itemize}
A study conducted by \cite{Bian:2008:FRF:1367497.1367561} found that users approach cQA forums more often to seek opinions and answers to complex questions than factoid based questions. In fact, these sites are successful primarily because they allow users to get fast and accurate answers to natural language questions, directly from a community. \\

\begin{table*}[htb]
\centering
\begin{tabular}{|l|l|l|}
\hline
\textbf{}                                                                     & \textbf{Community QA}                                                           & \textbf{QA}                                                                           \\ \hline
\textbf{Question Type}                                                        & \begin{tabular}[c]{@{}l@{}}Factoid Single \\ Sentence Questions\end{tabular}    & \begin{tabular}[c]{@{}l@{}}Multi Sentence \\ Questions\end{tabular}                   \\ \hline
\textbf{Source of Answers}                                                    & \begin{tabular}[c]{@{}l@{}}Extracted from \\ Documents in a corpus\end{tabular} & User generated                                                                        \\ \hline
\textbf{Quality of Answers}                                                   & Generally very high                                                             & \begin{tabular}[c]{@{}l@{}}Varies a lot, depending\\ on the community\end{tabular}    \\ \hline
\textbf{\begin{tabular}[c]{@{}l@{}}Availability of \\ Meta Data\end{tabular}} & None                                                                            & \begin{tabular}[c]{@{}l@{}}Best answer selection\\ and upvotes/downvotes\end{tabular} \\ \hline
\textbf{Time Lag}                                                             & Automatic and Immediate                                                         & \begin{tabular}[c]{@{}l@{}}Generally has to wait \\ For an answer\end{tabular}        \\ \hline
\end{tabular}
\caption{A comparison between cQA and QA}
\label{Comparison}
\end{table*}

Table \ref{Comparison} (taken from \cite{blooma2011research}) succinctly summarizes some of the key differences between a QA task and a cQA task. Most QA task deal with simple single sentence queries whose answers are simple facts. The questions are direct and rarely contain noise. On the other hand, cQA questions are seldom single sentenced, often with a lot of noise. (Eg, taken from yahoo answers, I have an exam tomorrow. But I don't know anything. Please recommend a tutorial for calculus ?? ). Moreover, in the former, answers are derived from a KB, while in the latter, the users respond and hence, the questions in a cQA task can be very open ended (Eg: How will Trump's presidency end ?). There is generally no handle on the quality of answers received on a community forum, but it also provides an access to other meta data like upvotes/ downvotes, answerer score etc. Consequently, cQA tasks allow for problems very different than the QA setting.     
%% -----END of cQA vs QA ------------------
\section{Tasks}
\subsection{Question Semantic Matching}
A big issue with such forums, with increasing number of questions, is that of question duplication, i.e if two questions are similar or not. Consider the following examples :
\begin{itemize}
\item \textit{What is the most populous state in India ?}
\item \textit{Which state in India has the highest population ?}
\end{itemize}
Both questions are essentially asking the same thing, i.e the answer for one can be relevant for the other and vice-versa. \\
Detecting such questions can be beneficial for a number of reasons: for one, it would lead to less redundancy; i.e if a person has answered the question once, he need not answer it again. Also, it would benefit the asker, for if there are numerous answers for the first question, and the asks asks the second question, then answers can be returned to him/her. \\
Here we discuss some approaches to the task
\begin{figure*}[!htb]
  \centering
  \subfloat[Embedding Based Methods]{\includegraphics[width=0.4\textwidth]{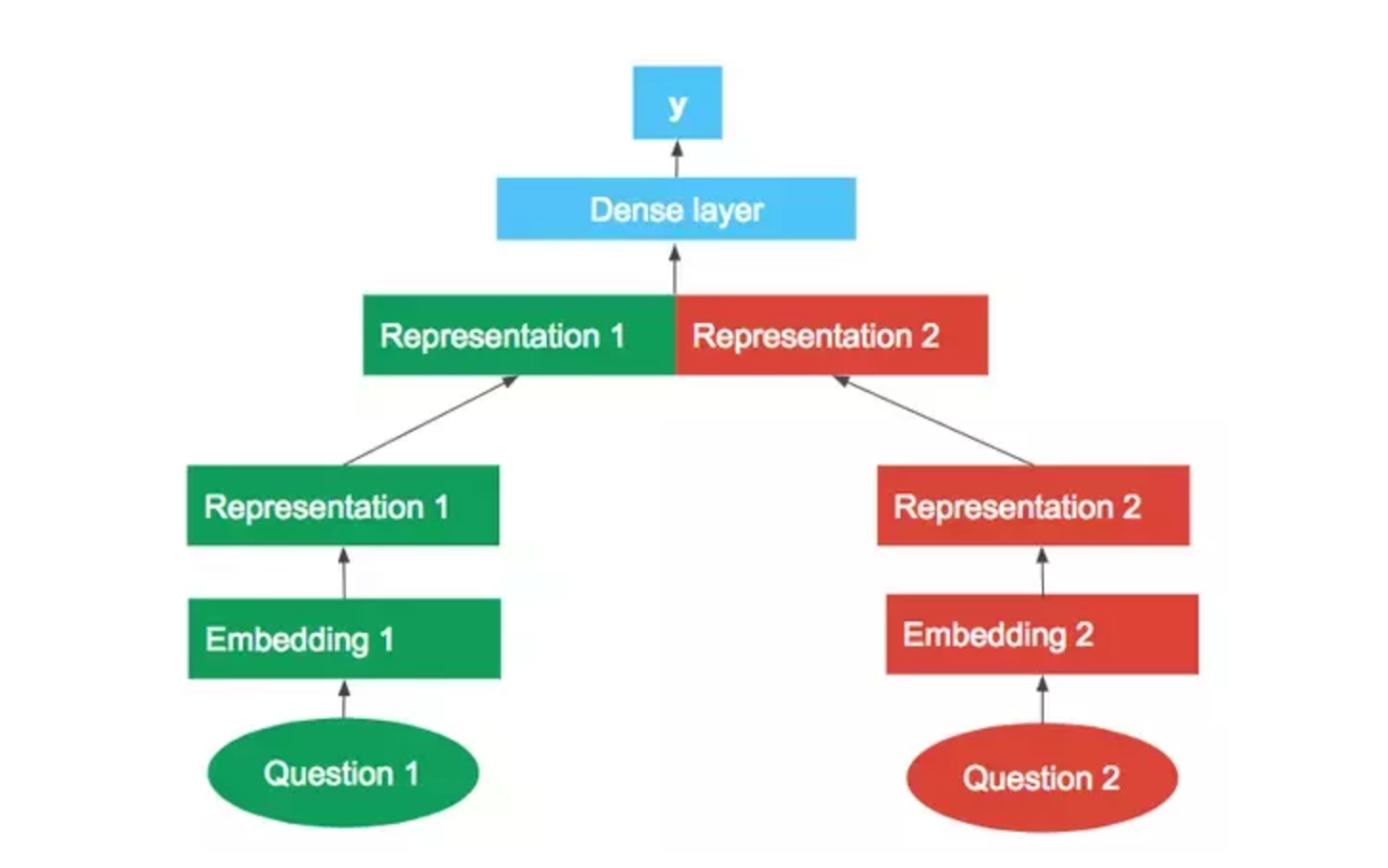}\label{fig:Emb_Met}}
  \hfill
  \subfloat[Neural Token Attention, taken from \cite{parikh2016decomposable}]{\includegraphics[width=0.6\textwidth]{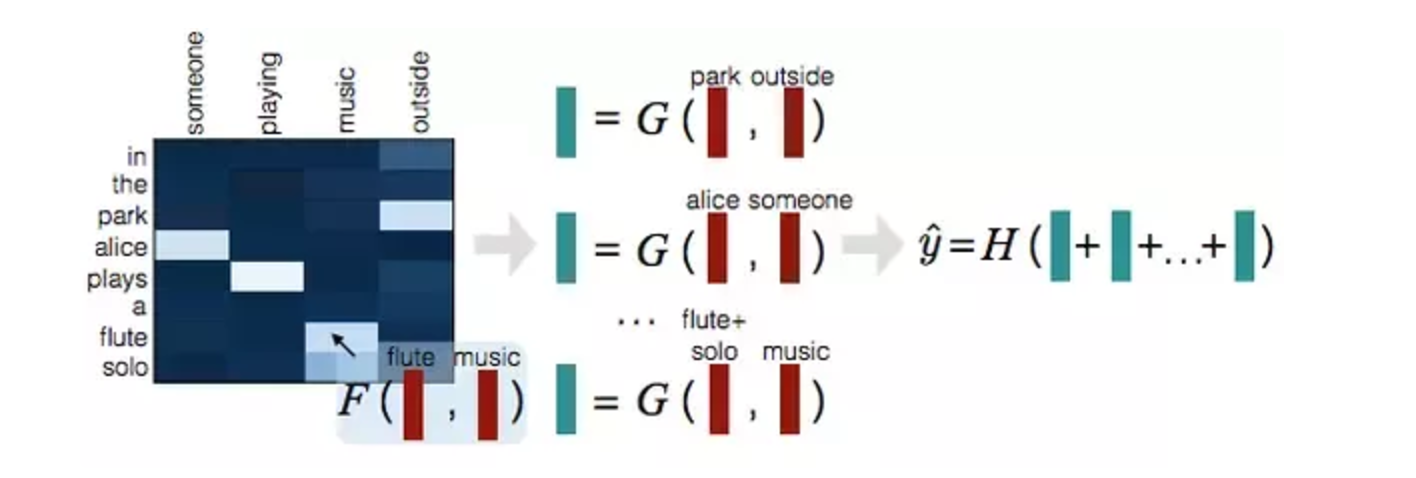}\label{fig:NTA}}
  \caption{The Embedding based methods }
\end{figure*}
\subsubsection{Okapi BM25 \cite{Robertson96okapiat}}

This method scores two question on the basis of their token similarity. The questions are treated as a bag of words, and a weighted matching (using Inverse Document Frequency) between the tokens is carried out for computing if two sentences are "close". The scoring function is defined as:
\begin{align}
\begin{split}
\label{eqn:Okapi}
 &Score(q^{1}, q^{2}) =  \sum_{i=1}^{|q^{1}|} IDF(q^{1}_{i}) freq(q^{1}_{i}, q^{2}) \\
 &freq(q^{1},q^{2}) = \frac{\#(q^{1}_{i}, q^{2}) (k_{1} + 1)}{ D^{r} } \\
 D^{r} &= \#(q^{1}_{i}, q^{2}) + k_{1}\cdot( 1-b + b\cdot \frac{(|q^{2}|)}{avg\_len(q^{2})} )
\end{split}
\end{align}
Where IDF is the inverse document frequency, $\#(w,Q)$ is the number of times w occurs in Q, avg\_len(Q) is the average length of Q and $k_{1}$ and b are tuned parameters. Since the scores are asymmetric, the final score is the average of $Score(q^{1},q^{2})$ and $Score(q^{2}, q^{1})$.

\subsubsection{TransLM \cite{xue2008retrieval}}
Given two questions, $q^{1}$ and $q^{2}$, a translation based language model is used to compute the probabilities $P(q^{1} | q^{2}) $ and $P(q^{2} | q^{1}) $, i.e the probability of generating a question, given the second. The score is taken as the average of the two probabilities. \\
The conditional probabilities consist of two parts: a smoothed version of the ML estimator for generating the words of the $q^{1}$, given $q^{2}$ (with the bag of words assumption), and the probability of generating $q^{1}$ given $q^{2}$, as a translation based model (i.e translating from $q^{2}$ to $q^{1}$). Specifically, the probability is computed as 
\begin{align}
\begin{split}
P(q^{1} | q^{2}) &= \prod_{w \in q^{1}} P(w|q^{2}) \\
P(w | q^{2}) &= \frac{|q^{2}|}{|q^{2}| + \lambda} \cdot P_{mx}(w|q^{2}) \\
 &+ \frac{\lambda}{|q^{2}| + \lambda}\cdot P_{ml}(w | C) \\
P_{mx}(w|q^{2}) &= (1 - \beta)P_{ml}(w|q^{2}) + \\
 &\beta \sum_{t\in q^{2}} P_{trans}(w|t) \cdot P_{ml}(t|q^{2}) 
\end{split}
\end{align}
Here, $P_{ml}(w|C)$ is the maximum likelihood estimate, computed as $\frac{\#(w,C)}{|C|}$, with $\#$ being the frequency. $\lambda$ is the smoothing factor, while $\beta$ controls the contribution of the $P_{ml}$ and $P_{trans}$. \\
$P_{trans}(w^{1}| w^{2})$, used traditionally in language translation models, computes the probability of generating word $w^{1}$ in a language, given $w^{2}$ in another language. Eg. given pairs of sentences $S = \{(e^{i},f^{i})\}_{i=1}^{N}$, the probability of translating an English word $e$ into a French word $f$ is given as :
\begin{align}
\begin{split}
P(f|e) &= \frac{1}{Z(e)} \sum_{i=1}^{N} c(f|e; e^{i},f^{i}) \\
c(f|e; e^{i},f^{i}) &= \frac{P(f|e)}{\sum_{w \in e^{i}}P(f|w)} \cdot \#(f,f^{i}) \#(e,e^{i})
\end{split}
\end{align}
Where Z(e) is the normalization constant. The values $P(f|e)$ are computed in an EM based method using the above equations. \cite{brown1993mathematics} show that this converges. \\
Now the problem of generating $q^{1}$ from $q^{2}$, is cast as a translation problem to compute $P_{trans}$.

\subsubsection{Word Embedding Based Methods}
This is a family of methods involving computing a real valued embedding of the two questions, and using a Multi Layer Perceptron. The embeddings may be the average/ sum of word vectors \cite{mikolov2013distributed}, trained on the data, or embeddings, learned in a joint model or can be obtained by passing the questions through an LSTM network \cite{hochreiter1997long}. (Figure \ref{fig:Emb_Met}\footnote{Taken from https://engineering.quora.com/Semantic-Question-Matching-with-Deep-Learning})

\subsubsection{Neural Network Attention with token alignment \cite{parikh2016decomposable}}
Given questions $q^{1} = [q^{1}_{1}, q^{1}_{2}, ..., q^{1}_{n}]$ and $q^{2} = [q^{2}_{1}, q^{2}_{2},..., q^{2}_{m}]$, using word embeddings for every token, new vectors $\hat{q}^{1} = [\hat{q_{1}}^{1}, \hat{q_{2}}^{1}, ..., \hat{q_{n}}^{1}]$ and $\hat{q}^{2} = [\hat{q_{2}}^{1}, \hat{q_{2}}^{2}, ..., \hat{q_{n}}^{2}]$ are obtained. An affine matrix L = $\sigma (\hat{q}^{1T} \cdot \hat{q}^{2} ) \in \mathcal{R}^{m x n}$ is generated. The affine matrix is then normalized row-wise and column wise (using softmax) to get the attention coefficients. The $j^{th}$ word in $q^{1}$ now is represented by $G[\hat{q_{j}}^{1};\hat{v}_{j}]$ where $\hat{v}_{j}$ is the attention weighted representation of $\hat{q}^{2}$, as defined by L and G is a non-linearity. Similarly, the new representations for each token of $q^{2}$ is obtained. The vector for the question is obtained as the sum of vectors for each token of the question, and then the two question representations are concatenated and a dense is applied to generate the prediction. (Figure \ref{fig:NTA})
\begin{figure*}[!thb]
  \centering
  \subfloat[QA-LSTM/CNN]{\includegraphics[width=0.4\textwidth]{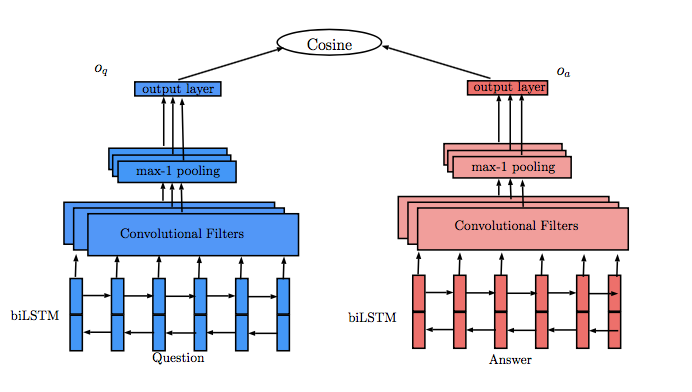}\label{fig:QA-LSTM_CNN}}
  \hfill
  \subfloat[QA-LSTM With Attention]{\includegraphics[width=0.6\textwidth]{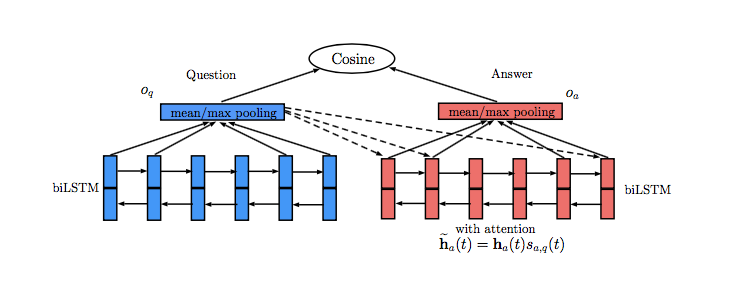}\label{fig:Attn_CNN}}
  \caption{The Embedding based methods, taken from \cite{tan2015lstm} }
\end{figure*}

\subsection{Question Answer Ranking and Retrieval}
Given the amount of traffic that a cQA site receives, the task of finding a good answer among the numerous answers posted is in itself an important one. Sometimes, the most upvoted answer may be noisy (Eg. a meme reply, or a sarcastic one liner often gets many upvotes, while not answering the question). Moreover, along with the question similarity task above, this can be used to find to new questions. This task is generally modeled as a task of answer selection : given a question q, and an answer pool ${a_{1}, ... a_{m}}$, we try and find the best candidate response. The candidate pool may or may not contain multiple gold labels. We discuss some methods to tackle the issue here.

\subsubsection{Okapi BM25 \cite{Robertson96okapiat}}
The method, described previously, can be used to score the responses, i.e answers with significant token overlap with the question would be scored higher. Since the token match of the questions and the answers seldom match, this method rarely performs very well.

\subsubsection{TransLM \cite{xue2008retrieval}}
The method, described previously, can be used to score responses. Specifically, given a question q, and an answer pool A, the problem of finding the best candidate answer can be modeled as the answer, which has the highest probability of generating the question. 
\begin{align}
a^{*} = argmax_{a \in A} P_{TransLM}(q|a)
\end{align}

\subsubsection{An embedding based CNN based method \cite{feng2015applying}}
This method, generates a question embedding and an answer embedding using a CNN network. Given question $q={q_{1}, ... q_{n}}$ and $a={a_{1}, .. ,a_{m}}$, the matrices $\hat{q} = [\hat{q}_{1}, ..\hat{q_{n}}] \in \mathcal{R}^{d \times n}$ and $\hat{a} = [\hat{a}_{1}, ..\hat{a_{m}}] \in \mathcal{R}^{d \times m}$ are generated, where d is the word embedding dimension. These word embeddings can be learned a priori using the Word2Vec model, or can be learned as a part of the model. A convolution filter of size m is applied along every dimension of the vector (generating a $\mathcal{R}^{d \times n - m + 1}$ for the question), following which a 1-max pooling layer (and an optional dense) is applied (generating a $\mathcal{R}^{d}$ vector). The CNN module is shared between the question and the answer.\\
Given a vector for the question, and an answer, the network is trained using a max margin method as described below:
\begin{align}
\begin{split}
\mathcal{L} = \sum_{(q,a) \in C} \sum_{(q,a') \in C'} max(0, \gamma - s(q,a) + s(q,a')) 
\end{split}
\end{align}
Where C is a set of questions with their correct answers, C' is a set of questions with an incorrect answer (obtained from negative sampling), $\gamma$ is a margin threshold, and s is a scoring function (cosine metric in this case).\\
A convolution network with filter size of 2, along with an average pooling was also used by \cite{DBLP:journals/corr/YuHBP14}. Instead of max-margin loss, their models, predicting 0 (for irrelevant responses) and 1 (for relevant responses) were trained by minimizing a log likelihood loss. They also use TF-IDF based counts as features.
\begin{figure*}
\centering
\includegraphics[width=0.6\textwidth]{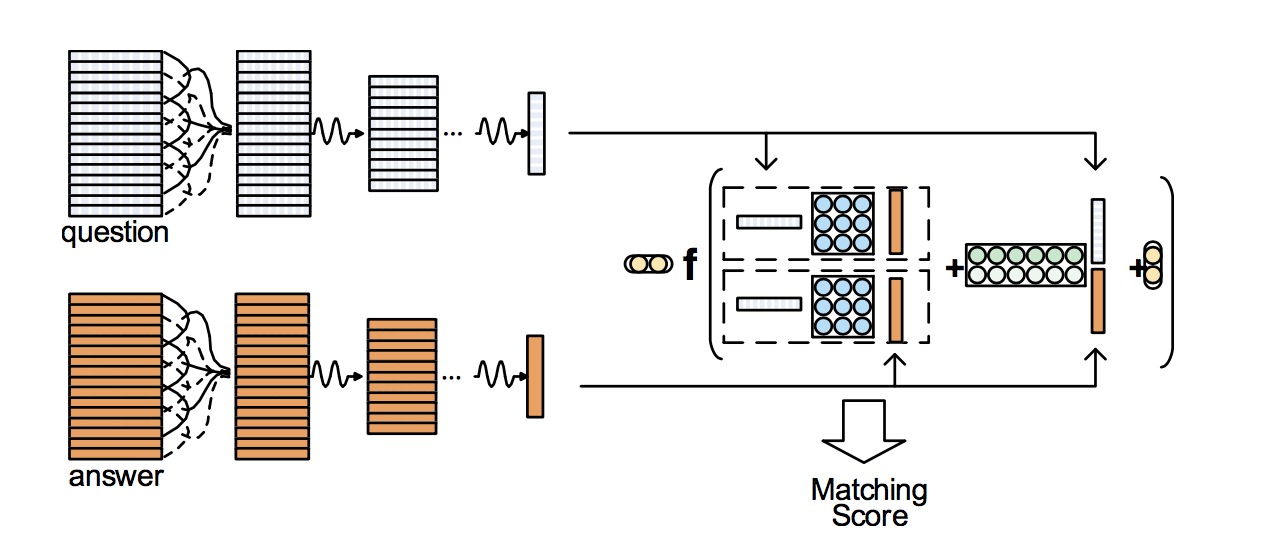}
\caption{The Architecture for the Neural Tensor Model, taken from \cite{qiu2015convolutional}}\label{fig:NTNModel}
\end{figure*}
\subsubsection{An Attention based CNN/LSTM Method \cite{tan2015lstm}}
Building on the previous work, the authors try and improve the model in two orthogonal directions. Instead of using just word embeddings, they pass the question and the answer through an Bi-LSTM layer. This allows them to encode the context. They then use a convolution layer and a max pooling layer. This allows them to capture long range dependencies better (the final state of the LSTM is somewhat limited by the dimension size for capturing the entire context) \\
In the attention based model (Figure \ref{fig:Attn_CNN}), following a max pooling of the question, the resultant vector is used to attend over the answer vectors. A max pool layer is then used over the attention weighted answer vectors, and the result is used as the embedding for the answer. This allows them to weigh the different words of the answer, based on the context, before the max pooling.\\
The final model presented combines both the ideas, using the CNN to generate the question embedding, using the question embedding to generate the attention weights for the answer, use the attention weighted answer as the input to the CNN module to generate the final answer embedding. The model is trained using the max margin loss, defined previously.

\subsubsection{A Deep CNN Method \cite{qiu2015convolutional}}
Building on the work of \cite{kalchbrenner2014convolutional}, the authors use deep convolutional networks to generate embeddings for given question and answer. A question $q = {q_{1} ... q_{m}}$ is first transformed into a matrix using word embeddings (learned as a part of the model), to get the input matrix $s = \mathcal{R}^{d \times l_{q}}$ where d is the dimension of the embedding layer, and $l_{q}$ is the length of q. Every row of this matrix is convolved with a different $\mathcal{R}^{m}$ filter, where m is the filter width (Hence, the number of convolving filters is $\mathcal{R}^{d \times m}$), and the resulting matrix formed is of dimensions $\mathcal{R}^{d \times (l_{q} - m + 1})$. \\
In order to make the convolutional network deeper, the model uses k-max pool layers. The k-max pool layer selects the k largest values along a dimension, and returns the subsequence without changing its relative order. The embedding dimensions hence are independent of the length of the question after the k-max pool (the dimension of the matrix is $\mathcal{R}^{d\times k}$ after the first k-max pool). The layer chooses the features contributing the maximum along a dimension, while preserving the order of the features. The value of k is chosen dynamically as $max(k_{top},\ceil*{\frac{D - d}{D} \cdot l_{q}} )$, where D is the maximum depth and d is the current depth. This process of convolutions followed by k-max pooling is done D times. A dense layer finally converts the $k_{top}$ vectors into a $\mathcal{R}^{n_{s}}$ vector ($v_{q}$). Similarly, a vector for the answer is obtained ($v_{a}$).  \\
The scoring function is also modified, to account for multiplicative as well as additive interactions between the vectors:
\begin{align}
\begin{split}
s(q,a) = U^{T}\sigma (v_{q}^{T} M^{[1:r]} v_{a} + V [v_{q} ; v_{a}] + b )
\end{split}
\end{align}
Where $M^{[1:r]} \in \mathcal{R}^{n_{s} \times n_{s} \times r}$ is a tensor capturing the multiplicative interactions (the bilinear tensor product $ v_{q}^{T} M^{[1:r]} v_{a}$ generates a vector $h \in \mathcal{R}^{r}$, and U, V and b are parameters. \\
This model can also be used for the question semantic matching task. Given a question q, finding the semantically closest query can be modeled as: 
\begin{align}
q^{*} = argmax_{(t,a) \in C} \alpha v_{q}^{T} v_{t} + (1 - \alpha)s(t,a)
\end{align}
Where C is the collection of question answer pairs and alpha controls the contribution of the question matching score and the question answer score (In case no answers are available, $\alpha = 1$.)
\section{Experiments}
The following datasets have been used for comparing the aforementioned methods.
\subsubsection{TREC QA Dataset for QA retrieval and Ranking}
The dataset, created by \cite{wang2007jeopardy}, contains a list of factoid question along with a list of answer sentences. The task is to rank candidate answers based on their relatedness to the question. Questions with all positive and all negative responses were removed for models by [\cite{feng2015applying} and \cite{tan2015lstm}], resulting in 1162 training questions, 65 dev questions and 68 test questions, while for the model by \cite{DBLP:journals/corr/YuHBP14}], the dataset (without any filtering) contained 1229 training, 82 dev and 100 test questions. The metrics used for comparing different models is the MAP (Mean Average Precision) and MRR (Mean Reciprocal Rank). 
\subsubsection{Yahoo! Answers Dataset}
The Dataset was generated by \cite{qiu2015convolutional}. Using the resolved questions for the Computer and Internet category in Yahoo! Answers, 312,000 question answer pairs were extracted. For the task of Question Answer retrieval, 10,000 positive pairs are used as the dev set, and 10,000 positive pairs are used as the test set. The rest are used for training, using 10 corrupted QA pairs (from negative sampling) per positive instance. \\
For the task of Question Semantic Matching, 5000 questions are sampled as the dev set, and 5000 as the test set. For ground truth generation, for every query, top-20 similar queries as denoted by a Vector Space Model are found (not containing the original query). The queries were then manually annotated by two annotators as similar/ relevant or irrelevant. A third annotator's decision was used as a tie breaker in cases of conflicts.
\subsubsection{Quora Dataset for Question Semantic Matching}
The dataset, released by Quora in 2017\footnote{https://data.quora.com/First-Quora-Dataset-Release-Question-Pairs}, has a collection of 404,351 sentence pairs, with associated labels : 149,306 positive examples (i.e semantically similar questions) and 255,045 negative examples. Alongside this, they are also running a Kaggle competition, with an additional test data set of 2,345,796 (unlabeled) sentences.
\section{Results}
Tables \ref{tab:TRECQA}, \ref{tab:Quora}, \ref{tab:QA_RET_Yahoo} and \ref{tab:QA_SEM_Yahoo} summarize the results of the aforementioned methods on the datasets.
\begin{table}[!htb]
\centering
\begin{tabular}{|l|l|l|}
\hline
\textbf{Models} & \textbf{MAP} & \textbf{MRR} \\ \hline
Bigram + Word Counts + CNN & 71.13 & 78.46 \\ \hline
Embedding + CNN + Max Pool & 71.06 & 79.98 \\ \hline
QA-LSTM & 68.19 & 76.52 \\ \hline
QA-LSTM/CNN & 70.61 & 81.04 \\ \hline
QA-LSTM with Attention & 68.96 & 78.49 \\ \hline
QA-LSTM/CNN with Attention & \textbf{72.79} & \textbf{82.40} \\ \hline
\end{tabular}
\caption{The results on the TREC QA Dataset}
\label{tab:TRECQA}
\end{table}
\begin{table}[!htb]
\centering
\begin{tabular}{|l|l|l|l|}
\hline
\textbf{Model} & \textbf{Prec} & \textbf{Rec} & \textbf{F Score} \\ \hline
Embedding (LSTM) & \textbf{88.0} & 86.0 & 87.0 \\ \hline
Neural Token Attn. & 81.0 & \textbf{95.0} & 87.0 \\ \hline
\end{tabular}
\caption{Results on the Quora dataset}
\label{tab:Quora}
\end{table}
\begin{table}[!htb]
\centering
\begin{tabular}{|l|l|}
\hline
\textbf{Models} & \textbf{P@1} \\ \hline
Okapi BM25 & 35.6 \\ \hline
TransLM & 48.5 \\ \hline
BOW Embeddings & 66.8 \\ \hline
CNN + MLP & 68.5 \\ \hline
CNTN & \textbf{70.7} \\ \hline
\end{tabular}
\caption{QA Retrieval on Yahoo! Answers}
\label{tab:QA_RET_Yahoo}
\end{table}
\begin{table}[!htb]
\centering
\begin{tabular}{|l|l|l|}
\hline
\textbf{Models} & \textbf{MAP} & \textbf{P@10} \\ \hline
Okapi & 32.5 & 22.9 \\ \hline
TransLM & 38.6 & 25.2 \\ \hline
NBOW & 39.2 & 26.8 \\ \hline
CNN + MLP& 41.8 & 27.4 \\ \hline
CNTN & \textbf{43.9} & \textbf{28.1} \\ \hline
\end{tabular}
\caption{Question Semantic Matching on Yahoo! Answers}
\label{tab:QA_SEM_Yahoo}
\end{table}
\section{Discussion}
As mentioned earlier, token matching methods work poorly for both the answer retrieval task and the question semantic matching task, because there is generally very little token overlap between questions and answers, while similar questions often are phrased such that they have different, but related words (this is termed as the lexical gap issue). Hence, it is not surprising that the performance of Okapi is very poor for both the tasks (Table \ref{tab:QA_RET_Yahoo} and Table \ref{tab:QA_SEM_Yahoo}). \\
Translation based methods, which model P(q|a) give an improvement the QA Retrieval task. This can be understood by the fact that certain words in the answer are indicative of the kind of questions that can be formed : Eg. century, hat-trick, One-Day, Record etc. in an answer are definitely indicative of a question about cricket. Since transitions between words in the answer and questions are modeled, one can expect such models to perform better. However, these models essentially treat the question and answer as a bag of words, and don't consider any semantics.  \\
Embedding based methods outperform the both the methods. This is to be expected, since embeddings do capture some semantic relations, which the other two cannot. We also see that simple word averaging is inferior to CNN/LSTM networks (Table \ref{tab:QA_RET_Yahoo} and Table \ref{tab:QA_SEM_Yahoo}). A reason for that could be that after averaging, the important words/contexts in the sentence could be diluted/lost and consequently, the representation obtained is poor. This could also explain why CNN's with max pooling layers also outperform general CNN's (Table \ref{tab:TRECQA}). We also see that the multiplicative interactions of question-answer embeddings help improve the performance (CNTN vs CNN + MLP) in both the retrieval and the matching task. \\
Note that the CNTN model is not specifically trained for the task of semantic matching. Hence, training a model specifically for the task may improve the performance, as is seen in Table \ref{tab:Quora}. \\
Table \ref{tab:TRECQA} provides a insight to the incremental value of CNN's and attention methods in the retrieval task. This somewhat supports the authors' claim that LSTM's hidden states fail to capture the entire context with increasing time-steps, and a convolution layer may help in retaining the important features across the sentence. Attending over the answer using a representation of the question also helps. An interesting experiment would be to see the models used for the Quora dataset (co-attention between the question - question tokens) perform fairly for the TREC-QA task. \\
Finally, the convolution methods for the Attention based CNN and the Deep CNN (and the embedding based CNN method) methods are somewhat different. Deep CNN uses d filters, convolving across each dimension separately, while the attention based CNN uses the following method :
\begin{align}
o_{F}(t) = tanh ( (\sum_{i=0}^{m-1} h(t+i)^{T} F(i)) + b )
\end{align}
The dimensions in the resultant matrix comes from different such filters. It can be shown that the first convolution method is a special case of the second, by setting the appropriate weights to zero. A model, combining the convolution method of the Attention based CNN with the depth using k-max pooling would be interesting.  
\section{Future Work and Conclusion}
This survey aims at looking at some of the tasks in cQA and the methods used to tackle them. Some other tasks in this domain, that have not been covered here are Expert Finding (finding the experts in a Community Forum) and finding comment significance (i.e identifying the relevant comments to a question in a comment thread - the SemEval-2015 task \cite{alessandromoschitti2015semeval}). Other methods used, that have not been covered here are Tree Kernel based semantic question matching (\cite{wang2009syntactic}) and using syntactic features (\cite{carmel2014improving}). These methods use a lot of feature engineering and syntactic features, but they they suffer from errors of their individual components (POS Taggers, Dependency Parsers etc.) \\ 
An issue in this domain is the lack of a single dataset to compare different models. Many authors use the Yahoo! Answers dataset but the data collection methods  and consequently the dataset is not fixed. This makes it very difficult to compare models across papers. While many authors evaluate their models on the TREC-QA dataset, the dataset contains only factoid questions, and consequently is not a very good representative of community forum posts \cite{Bian:2008:FRF:1367497.1367561}.  An area of future work could be testing the different models on some common datasets, and carrying out an extensive error analysis to compare their relative merits and demerits.
%\bibliography{acl2017}
\bibliography{acl2017}
\bibliographystyle{acl_natbib}

\end{document}